\documentclass[aps,pra,reprint,superscriptaddress,showpacs]{revtex4-1}

\usepackage{amsmath,amssymb,amsfonts}
\usepackage[english]{babel}
\usepackage{graphicx}
\usepackage{bm}

\begin{document}

\title{Propagation and perfect transmission in three-waveguide axially varying couplers}

\author{B. M. Rodr\'{\i}guez-Lara}
\affiliation{Instituto Nacional de Astrof\'{i}sica, \'{O}ptica y Electr\'{o}nica \\ Calle Luis Enrique Erro No. 1, Sta. Ma. Tonantzintla, Pue. CP 72840, M\'{e}xico}
\email{bmlara@inaoep.mx}

\author{H. M. Moya-Cessa}
\affiliation{Instituto Nacional de Astrof\'{i}sica, \'{O}ptica y Electr\'{o}nica \\ Calle Luis Enrique Erro No. 1, Sta. Ma. Tonantzintla, Pue. CP 72840, M\'{e}xico}

\author{D. N. Christodoulides}
\affiliation{CREOL/College of Optics and Photonics, University of Central Florida, Orlando, Florida, USA}

\begin{abstract}
We study a class of three-waveguide axially varying structures whose dynamics are described by the $su(3)$ algebra.
Their analytic propagator can be found based on the corresponding Lie group generators.
In particular, we show that the field propagator corresponding to three-waveguide structures that have arbitrarily varying coupling coefficients and identical refractive indices is associated with the orbital angular momentum algebra. 
The conditions necessary to achieve perfect transmission from the first to the last waveguide element are obtained and particular cases are elucidated analytically.
\end{abstract}


\maketitle

\section{Introduction} \label{sec:S1}

Waveguide  couplers represent indispensable  elements in integrated optics. 
Such structures find numerous applications in several areas of optics, especially in those pertaining to the switching and routing of light \cite{Marcuse1973,Hunsperger2009}.
Typically waveguide couplers are theoretically analyzed using coupled-mode theory \cite{Jones1965p261,Marcuse1973p817,Yariv1973p919}. 
Along these lines, certain classes of directional couplers amenable to analytic solutions have been explored by assuming adiabatic changes in their coupling parameters \cite{Milton1975p1207,Schneider2001p129,Huang2006p056606,Paspalakis2006p30,Longhi2007p201101,Lahini2008p193901,Salandrino2009p4524}.
In such adiabatic directional couplers the local eigenstates remain invariant under slow perturbations in the device characteristics. 
However, the restriction of having slow, gradual changes in the parameter space of these devices constrains the spatial configurations where directional adiabatic couplers can be implemented.

In this report we study a class of coupled, axially varying, three-waveguide structures whose dynamics are described by the $su(3)$ algebra. 
The analytic propagator of these systems can be constructed from the corresponding Lie group generators. 
Specifically, we are interested in new classes of planar three-waveguide directional structures with arbitrary varying coupling coefficients and we focus on the construction of a field propagator for a particular sub-class of three-waveguide structures described by the orbital angular momentum algebra.  
Our class of three-waveguide couplers is not restricted by slow gradual changes in the characteristics of the photonic crystal and, thus, may extend the spatial configurations for the design of these devices.
The transfer of classical and quantum states of light encoding information is of paramount importance in scalable optical and quantum processing \cite{PerezLeija2013p012309}; for this reason we identify the conditions necessary to synthesize a three-waveguide coupler that provides perfect transmission from the first to the last waveguide element.
These results are elucidated via relevant examples.

\section{Underlying model and its solution} \label{sec:S2}

We here consider a three-core waveguide structure with arbitrary axially varying nearest neighbor couplings and refractive indices.  
Using coupled mode theory, this system can be described by the following set of differential equations 
 \cite{Jones1965p261,Marcuse1973p817,Yariv1973p919, Xia2013},
\begin{eqnarray}
\frac{d \mathbf{E}(z)}{d z} =  \imath \left(
\begin{array}{ccc}
n_{1}(z) & \alpha(z) & 0 \\
\alpha(z) & n_{2}(z) & \beta(z) \\
0 & \beta(z) & n_{3}(z)
\end{array}
\right) \mathbf{E}(z), 
\end{eqnarray}
where the modal field amplitude corresponding to each waveguide, $\mathcal{E}_{j}$, is contained in the vector $\mathbf{E}(z) = \left( \mathcal{E}_{1}(z), \mathcal{E}_{2}(z), \mathcal{E}_{3}(z) \right)$.
The nearest neighbor couplings are given by the well behaved real functions $\alpha(z)$ and $\beta(z)$ associated with the coupling coefficient between the first and second waveguide and that between the second and third waveguide, respectively. 
The effective refractive index $n_{j}(z)$ of each waveguide is also a well behaved real function.
Throughout this work we also assume that in addition $\sum_{j} n_{j}(z) = n$; i.e., the sum of the three indices is a constant.
Under such conditions, it is straightforward to write the differential system in terms of Gell-Mann-Ne'eman matrices \cite{GellMann1961,Neeman1961p222},
\begin{eqnarray}
\frac{d \mathbf{E}(z)}{d z} = i \left[\alpha(z) \bm{\lambda}_{1} +  \beta(z) \bm{\lambda}_{6} + \gamma(z) \bm{\lambda}_{3} + \delta(z) \bm{\lambda}_{8} \right] \mathbf{E}(z),   \label{eq:SchEq}
\end{eqnarray}
where here we have omitted a unitary matrix term proportional to the constant $n/3$ that only introduces an overall phase factor.
The new parameter functions are defined as $\gamma(z) = \left[ n - 2 n_{2}(z) - n_{3}(z) \right]/2$ and $\delta(z) = \left[ n - 3 n_{3}(z)\right]/(2 \sqrt{3})$.

Equation \eqref{eq:SchEq} is a Schr\"odinger equation up to an overall phase factor.
By following standard quantum mechanical techniques, cf. \cite{Dattoli1991p1247} and references therein, one can show that the propagator for the field amplitudes, $\mathbf{E}(z)= \mathbf{U}(z) \mathbf{E}(0)$, is given by  
\begin{eqnarray}
U(z) = \left( \begin{array}{ccc} 
 f_{1}(z) & -f_{2}(z) & -f_{3}(z) \\
-g_{1}(z) &  g_{2}(z) &  g_{3}(z) \\
-h_{1}(z) &  h_{2}(z) &  h_{3}(z)
\end{array}\right)
\end{eqnarray}
where 
\begin{eqnarray}
\frac{d}{dz} \left( \begin{array}{c}
f_{j}(z) \\ g_{j}(z) \\ h_{j}(z)
\end{array} \right) = i \left( \begin{array}{ccc} 
 n_{1}(z) & -\alpha(z) & 0 \\
 -\alpha(z) &  n_{2}(z) & \beta(z) \\
0 &  \beta(z) &  n_{3}(z)
\end{array}\right) \left( \begin{array}{c}
f_{j}(z) \\ g_{j}(z) \\ h_{j}(z)
\end{array} \right) \label{eq:AuxDiffSet}
\end{eqnarray}
with $j=1,2$, and $3$ and initial conditions given by $f_{1}(0)=g_{2}(0)=h_{3}(0)=1$ and zero for the rest, $f_{2}(0)=f_{3}(0)=g_{1}(0)=g_{3}(0)=h_{1}(0)=h_{2}(0)=0$.
At this point, we only need to find the forms of the complex functions for each specific case.
For the sake of simplicity we will here focus on the subcase where the three waveguides exhibit indentical refractive indices, which implies that $n_{j}= n/3$; in doing so, one must be careful in choosing appropriate coupling symmetries that keep the length of all waveguides approximately equal (within a small fraction of a wavelength) so as to avoid phase accumulation effects. 
In this case we can write the differential set describing the system in the following way:
\begin{eqnarray}
\frac{d \mathbf{E}(t)}{d t} = \imath \left[ \bm{\lambda}_{1} + \eta(t) \bm{\lambda}_{6}
\right] \mathbf{E}(t), \label{eq:DifSetRed}
\end{eqnarray}
where we have defined a coupling ratio $\eta(z) = \beta(z)/\alpha(z)$ and made the variable change $\alpha(z) = \partial t / \partial z$.
We can take advantage of the fact that the Gell-Mann matrices $\{ \lambda_{1}, \lambda_{5}, \lambda_{6} \}$ form an orbital angular momentum, $SU(2)$, which is a sub-group of $SU(3)$.
Then the propagator can be written as follows:
\begin{eqnarray}
\bm{U}(t) = e^{i f(t) \bm{S}_{+} } e^{i g(t) \bm{\lambda}_{6} } e^{i h(t) \bm{S}_{-} },
\end{eqnarray}
with $\bm{S}_{\pm} = \bm{\lambda}_{5} \pm i \bm{\lambda}_{1}$ and 
\begin{eqnarray}
f(t) &=&  2 i \frac{ \nu^{\prime}(t)}{\nu(t)} - \eta(t),  \label{eq:Eqft}\\
g(t) &=& 2 i \left[ \ln \nu(t) - \ln \nu(0) \right], \label{eq:Eqgt}\\
h(t) &=& \frac{\nu^{2}(0)}{2}  \int_{0}^{t} dx~ 
\frac{1}{\nu^{2}(x)}.\label{eq:Eqht}
\end{eqnarray}
where we have introduced an auxiliary function $\nu(t)$ that obeys the following second order differential equation
\begin{eqnarray}
\nu ''(t)+\frac{1}{4} \nu (t) \left[ 2 i \eta '(t)+\eta^2 (t) + 1 \right]&=&0,  \label{eq:DiffEq}
\end{eqnarray}
provided that $f(0)=0$. 
In general, Eq. \eqref{eq:DiffEq} is not known to exhibit analytical solutions for any arbitrary coupling ratio $\eta(z)$.
Nevertheless, as we will see, this problem can be solved in closed form for particular coupling ratios. 

\begin{figure}
\includegraphics[scale=1]{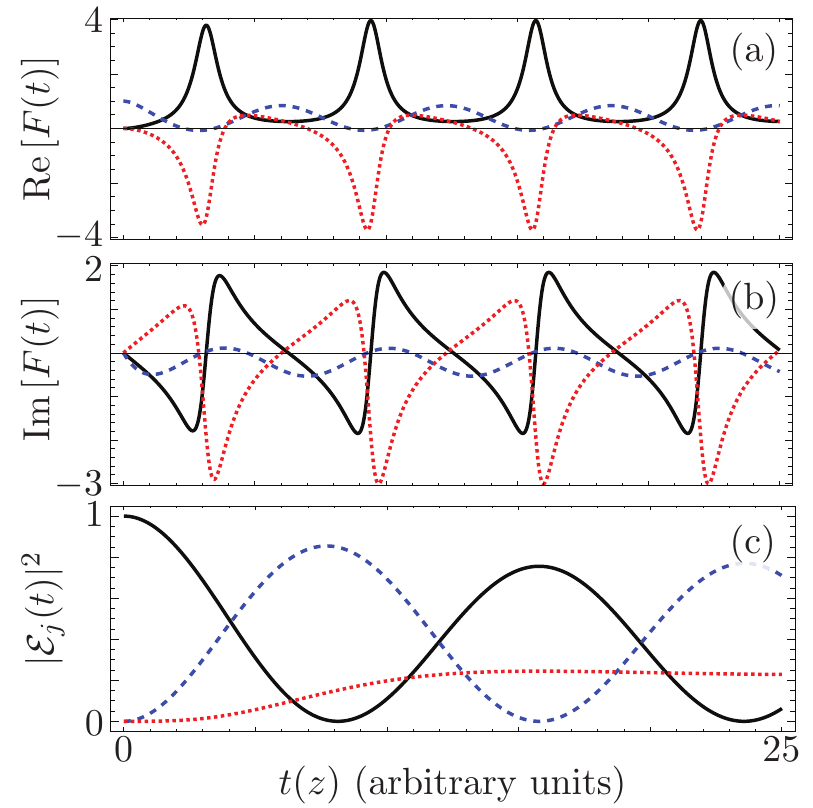}
\caption {(Color online) The (a) real and (b) imaginary parts of $F(t)=$ $f(t)$ (solid black line), $e^{- i g(t)}$ (dashed blue line) and $h(t)$ (dotted red line) and (c)  the intensity at the first (solid black line), second (dashed blue line) and third (dotted red line) waveguides for an initial field amplitude impinging the first waveguide of a device with the exponential coupling ratio $\eta(t)= e^{-t}$.}  \label{fig:Fig1}
\end{figure}

Let us consider for example the case of exponential coupling ratios $\eta(t) = c e^{dt}$ with $c,d \in \mathbb{R}$.
We introduce an extra change of variable, $\nu(t) = e^{\frac{i}{2d} \left(d t-c e^{d t}\right)} u(t)$,
such that the differential equation for the auxiliary function \eqref{eq:DiffEq} becomes:
\begin{eqnarray}
u''(t) + \left(i-i c e^{d t}\right) u'(t)+\frac{1}{2} c e^{d t} u(t) = 0.
\end{eqnarray}
Further manipulation provides a solution of the form, 
\begin{eqnarray}
u(t) &=& c_1 ~_1F_1 \left(\frac{i}{2 d},\frac{d+i}{d},\frac{i c e^{d t}}{d}\right)  \nonumber \\ 
&& +  c_2 ~ d^{i/d} c^{-\frac{i}{d}} e^{-\frac{\pi }{2 d}-i t} ~_1F_1\left(-\frac{i}{2 d},\frac{d-i}{d},\frac{i c e^{d t}}{d}\right), 
\end{eqnarray}
where $_{1}F_{1}(\alpha,\beta,x)$ is the confluent  hypergeometric function of Kummer \cite{Kummer1837p228,Lebedev1965,Abramowitz1970} and the constants $c_{j}$ are to be chosen in order to satisfy the conditions $f(0)=g(0)=h(0)=0$. 
The rest of the solution can be calculated via Eqs. \eqref{eq:Eqft}-\eqref{eq:Eqht}.
Again, the functions $f(t)$, $g(t)$, and $h(t)$ are also given in terms of confluent hypergeometric and exponential functions.
Figures \ref{fig:Fig1}(a) and \ref{fig:Fig1}(b) show the real and imaginary parts of these functions, respectively, as well as the intensity at each waveguide for an initial field impinging on the first waveguide,  Fig. \ref{fig:Fig1}(c), for the coupling ratio $\eta(t) = e^{-t}$. 

\section{Perfect transmission in the $SU(2)$ model} \label{sec:S3}

Let us consider light impinging on the three-waveguide coupler at the first waveguide site, $\mathbf{E}(0)=(1,0,0)$.
The field amplitudes through this structure are given by the following expressions:
\begin{eqnarray}
\mathcal{E}_{1}(t) &=&  1-2 f(t) e^{-i g(t)} h(t) \label{eq:a1z}\\
\mathcal{E}_{2}(t) &=&  -f(t)+\left[f^2(t)+1\right] e^{-\imath g(t)} h(t) \label{eq:a2z} \\
\mathcal{E}_{1}(t) &=&  -f(t)+\left[f^2(t)-1\right] e^{-\imath g(t)} h(t) . \label{eq:a3z}
\end{eqnarray}
In order to obtain perfect transmission from the first to the last waveguide, we have to search for a specific length $\tau$ at which $\mathbf{E}(\tau)=(0,0, e^{i \phi})$.
This is possible when
\begin{eqnarray}
f(\tau) &=& \pm 1, \quad \phi = 0,\pi , \\
h(\tau) &=& \pm \frac{1}{2} e^{\imath  g(\tau)} .
\end{eqnarray}
Of course, not all the cases of coupling ratios $\eta(t)$ will satisfy these requirements and, typically, it is complicated to find out if they are fulfilled or not.

\subsection{Constant coupling ratio: $\eta(z) = c $ with $c \in \mathbb{R}$} \label{sec:S2a}

\begin{figure}
\includegraphics[scale=1]{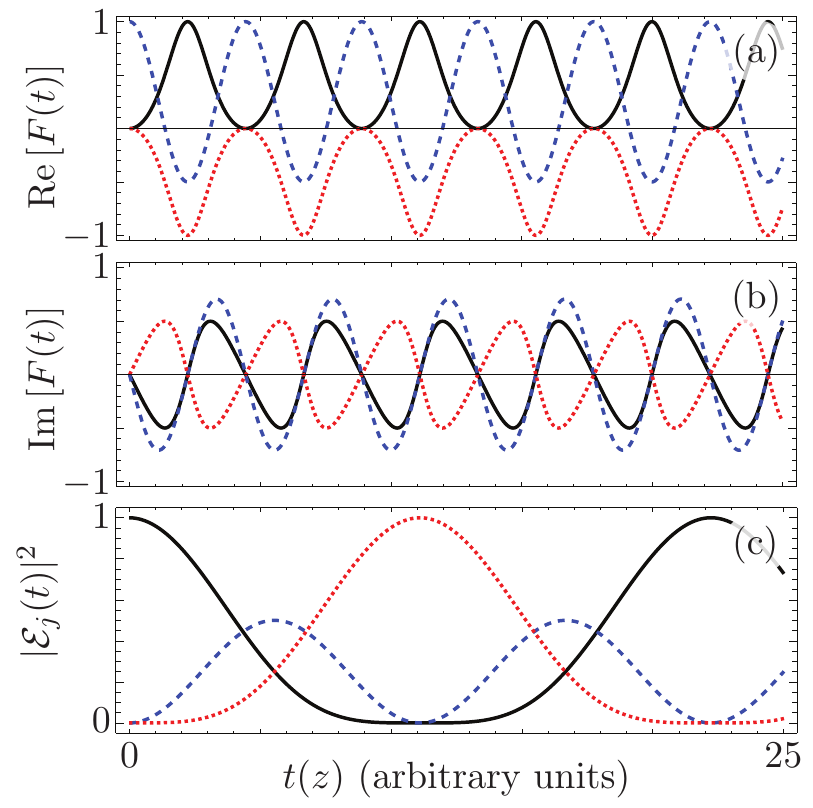}
\caption {(Color online) The (a) real and (b) imaginary parts of $F(t)=$ $f(t)$ (solid black line), $e^{- i g(t)}$ (dashed blue line) and $h(t)$ (dotted red line) and (c) the intensity at the first (solid black line), second (dashed blue line) and third (dotted red line) waveguides for an initial field amplitude impinging the first waveguide of a device with the constant coupling ratio $\eta(t)= 1$.}  \label{fig:Fig2}
\end{figure}

To demonstrate the versatility of our approach, we first use it in the  case where the coupling ratio $\eta(z)$ is constant. 
In this case it is straightforward to show that the auxiliary function is given by
\begin{eqnarray}
\nu(t) =  c_{+} e^{\frac{i}{2}  \sqrt{1+c^2} t} + c_{-} e^{\frac{i}{2}  \sqrt{1+c^2} t },
\end{eqnarray}
where the coefficients $c_{\pm}$ are to be determined form the initial conditions. 
Thus, the Eqs. \eqref{eq:Eqft}-\eqref{eq:Eqht} yield
\begin{eqnarray} 
f(t)&=& \frac{1}{c+i \sqrt{a} \cot \left(\frac{1}{2} \sqrt{a} t\right)}, \\
g(t)&=& i \ln \left[ \frac{\sqrt{a} \cos \left(\sqrt{a} t\right)- i c \sin \left(\sqrt{a} t\right)}{\sqrt{a} \left[ 1+f^2(t)\right]}  \right]\\
h(t)&=& -f(t),
\end{eqnarray}
with $a= 1 + c^2$.
Note that for this particular case it is simpler just to calculate the matrix exponential $e^{i \left( \hat{S}_{y} + c \hat{S}_{z} \right) z}$ required in the solution,
\begin{eqnarray}
\bm{U}(t) &=&  \left(
\begin{array}{ccc}
 \frac{c^2+\cos \left(\sqrt{a} t \right)}{a} & \frac{i \sin \left(\sqrt{a} t \right)}{\sqrt{a}} & \frac{c \left(\cos \left(\sqrt{a} t \right)-1\right)}{a} \\
 \frac{i \sin \left(\sqrt{a} t \right)}{\sqrt{a}} & \cos \left(\sqrt{a} t \right) & \frac{i c \sin \left(\sqrt{a} t \right)}{\sqrt{a}} \\
 \frac{c \left(\cos \left(\sqrt{a} t \right)-1\right)}{a} & \frac{i c \sin \left(\sqrt{a} t \right)}{\sqrt{a}} & \frac{\cos \left(\sqrt{a} t \right)  + c^2}{a} \\
\end{array}
\right). \label{eq:UtConstant}
\end{eqnarray}
It is straightforward to realize that the case of uniformly coupled waveguides \cite{Jones1965p261} is included in this class and that the auxiliary functions oscillate periodically.
Figure \ref{fig:Fig2} shows the real and imaginary parts of these functions for the case $\eta(t)=1$.
This case supports perfect transmission from the first to the third waveguide as shown in Fig. \ref{fig:Fig2}(c).

\subsection{Linear coupling ratio: $\eta(t) = c t$ with $c \in \mathbb{R}$} \label{sec:S2b}

\begin{figure}
\includegraphics[scale=1]{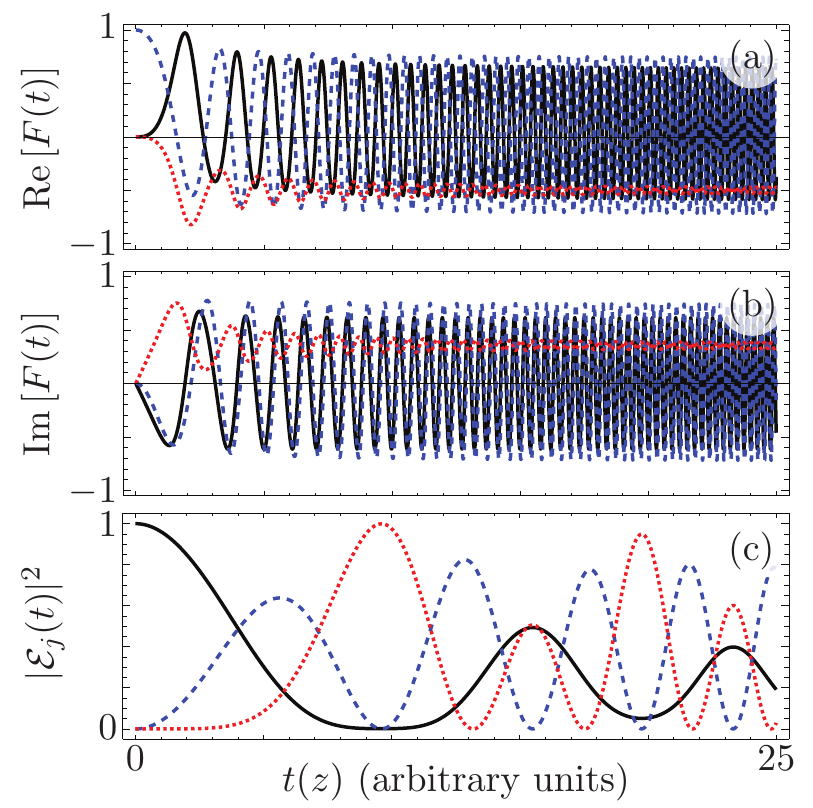}
\caption {(Color online) The (a) real and (b) imaginary parts of $F(t)=$ $f(t)$ (solid black line), $e^{- i g(t)}$ (dashed blue line) and $h(t)$ (dotted red line)  and (c) the intensity at the first (solid black line), second (dashed blue line) and third (dotted red line) waveguides for an initial field amplitude impinging the first waveguide of a device with the linear coupling ratio $\eta(t)= t$.}  \label{fig:Fig3}
\end{figure}

The case of a linear coupling ratio, $\eta(t) = c t$ with $c \in \mathbb{R}$, can be solved analytically by the auxiliary function \begin{eqnarray}
\nu(t) &=& c_{e} e^{- i \alpha t^2 /4} ~_{1}F_{1}\left( \frac{ i}{8 c}, \frac{1}{2}, \frac{ic}{2} t^2 \right).
\end{eqnarray}
Note that  the conditions $\nu(0) = c_{e}$ and $\lim_{t \rightarrow \infty} \nu(t) = 0$ allows us to write the rest of the functions in the following form:
\begin{eqnarray}
f(t) &=& - \frac{ i t}{2}~ \frac{ ~_{1}F_{1}\left(1 + \frac{ i}{8 c},\frac{ 3}{2}, \frac{ic}{2} t^2 \right)}{~_{1}F_{1}\left( \frac{ i}{8 c}, \frac{1}{2}, \frac{ic}{2} t^2 \right)}, \\
g(t) &=& 2 i \ln \left[  c_{e} e^{- i \alpha t^2 /4} ~_{1}F_{1}\left( \frac{ i}{8 c}, \frac{1}{2}, \frac{ic}{2} t^2 \right) \right], \\
h(t) &=&  c_{e} \int_{0}^{t} dx ~e^{i \alpha x^2 / 2} ~_{1}F_{1}^{-2}\left( \frac{ i}{8 c}, \frac{1}{2}, \frac{ic}{2} x^2 \right).
\end{eqnarray}
The function $h(t)$ can be seen as a generalization of Fresnel integrals leading to a series of confluent hypergeometric or incomplete gamma functions via expansion of the confluent hypergeometric function into a power series \cite{Abramowitz1970}. 
These functions fulfill the initial condition $f(0)=g(0)=h(0)=0$. 
Figure \ref{fig:Fig3} shows the real and imaginary parts of these functions for the coupling ratio case $\eta(t) = t$ leading to an auxiliary coefficient $c_{e}=1$. 
Notice the asymptotic localization of the function $e^{- i g(t)}$ as expected from the asymptotic behavior of the confluent hypergeometric function.
Clearly, this case supports perfect transmission from site 1 to 3 as shown in Fig. \ref{fig:Fig3}(c).

\subsection{Optical analog of STIRAP}

\begin{figure}
\includegraphics[scale=1]{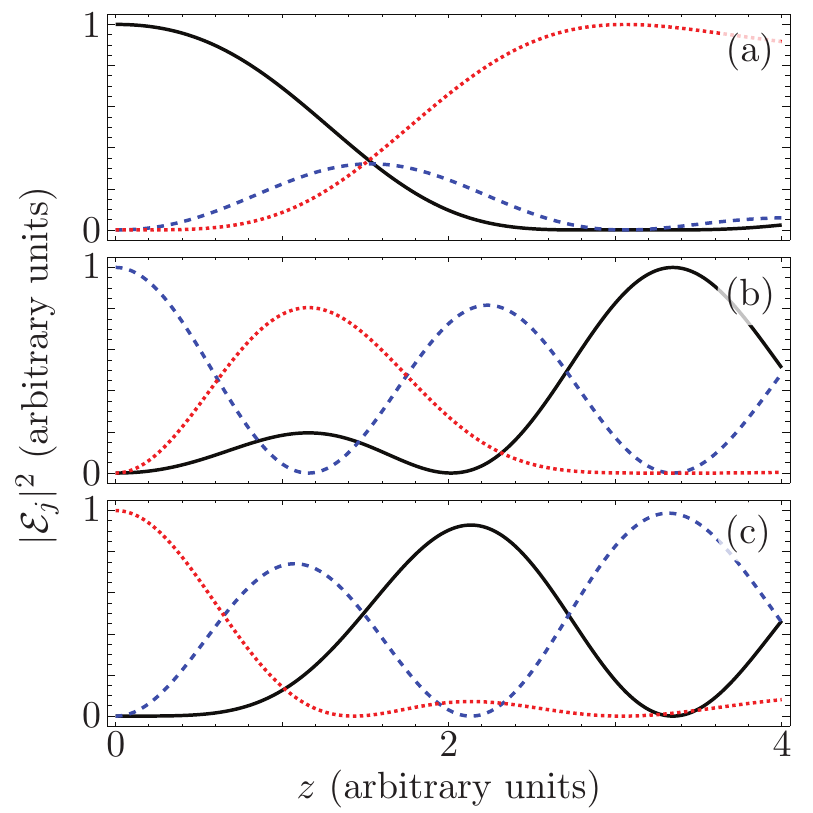}
\caption {(Color online) Intensity propagation through the first (solid black line), second (dashed blue line) and third (dotted red line) waveguides showing one-directional coupling in a system with parameters $c=2.5$ and $ \zeta = 3$. The initial field is impinging the (a) first, (b) second, and (c) third waveguide. }  \label{fig:Fig4}
\end{figure}

Another case included in the $SU(2)$ model providing perfect transmission are the optical analogs of stimulated Raman adiabatic passage (STIRAP) corresponding to parameter values: $\alpha(z) = c e^{-(z-\zeta)^2 / \zeta^2 }$ and $\beta(z) = c e^{-z^2 / \zeta^2 }$ where $c, \zeta \in \mathbb{R}$.
Under these restrictions, our optical coupler model is equivalent, up to a constant phase factor, to an atomic three-level delta system driven by two classical fields as presented in \cite{Mitra2003p043409}; for a review of STIRAP cf. \cite{Bergmann1998p1003,Vitanov2001p763}.
In this case the auxiliary variable $t(z)$ leads to the error function and beyond that the problem can only be tackled via the method of Frobenius.
The results obtained are in the form of power series and recurrence relations for the coefficients.
Unitary integration techniques \cite{Rau1998p4785} may provide a more compact form for the propagator  but such an approach is beyond the scope of this report.
Figure \ref{fig:Fig4} shows the intensity at each waveguide under different initial conditions and demonstrates the one-directional coupling characteristics of this system.

\section{Conclusions} \label{sec:S6}

We have shown that it is possible to solve the light evolution equations in a nonadiabatic axially varying three-core waveguide coupler involving non-identical waveguides, in terms of the Lie group generators of $su(3)$.
We focused our attention on a reduced class of structures having identical waveguides where the dynamics is dictated by the orbital angular momentum algebra, $su(2)$.
The advantage of this specific method resides in the fact that one has to solve only a second order differential equation involving the three-waveguide coupling ratio and its derivative in order to determine the three auxiliary functions required in the orbital angular momentum rotations.
As an example we consider the case of exponential coupling ratios.
For perfect transmission, we found the conditions needed for these three auxiliary functions to allow complete transmission from the first to the last waveguide in the coupler at a given propagation distance.
We approached the simplest class of constant and linear coupling ratios as a practical example of such systems allowing perfect transmission; in these cases the solution to the differential equation can be written in closed form in terms of trigonometric and hypergeometric functions.
The constant coupling ratio case leads to periodically oscillating fields in the coupler channels  and perfect transmission for a well defined set of parameters as expected.
The linear coupling ratio case provides close to perfect transmission with a larger set of parameters.
Our analytic results point to a class of coupling ratios in photonic three-waveguide couplers that provides directional coupling with almost perfect transmission from the first to the last waveguide.
This fact allows for the design of directional couplers in a variety of configurations.
Finally, we studied the optical analog of STIRAP corresponding to the $SU(2)$ model and found its solution as a set of recurrence relations and power series. 
Alternative methods, such as unitary integration techniques, may provide closed form solutions.

\begin{acknowledgments}
BMRL gratefully acknowledges valuable comments and suggestions from Francisco Soto-Eguibar.
\end{acknowledgments}


\end{document}